\documentclass[preprint,showpacs,preprintnumbers,amsmath,amssymb]{revtex4}

\newcommand\rf[1]{(\ref{eq:#1})}
\newcommand\lab[1]{\label{eq:#1}}

\newcommand\br{\begin{eqnarray}}
\newcommand\er{\end{eqnarray}}
\newcommand\be{\begin{equation}}
\newcommand\ee{\end{equation}}

\newcommand\lb{\lbrack}
\newcommand\rb{\rbrack}

\newcommand\bc{\begin{center}}
\newcommand\ec{\end{center}}

\newcommand\pa{\partial}

\newcommand{\ct}[1]{\cite{#1}}
\newcommand{\bib}[1]{\bibitem{#1}}
\newcommand\PRL[3]{\textsl{Phys. Rev. Lett.} \textbf{#1}, #3 (#2)}

\begin{document}
\preprint{arxiv:[hep-th]}

\title{Photon and Axion Splitting in an Inhomogeneous Magnetic Field }

\author{E.I. Guendelman}%
\email{guendel@bgumail.bgu.ac.il}
\affiliation{%
Department of Physics, Ben-Gurion University of the Negev \\
P.O.Box 653, IL-84105 ~Beer-Sheva, Israel
}%

\begin{abstract}
The axion photon system in an external magnetic field, when the direction of propagation of axions
and photons is orthogonal to the direction of the external magnetic field, displays a continuous axion-photon 
duality symmetry in the limit the axion mass is neglected. The conservation law that follow in this effective 
$2+1$ dimensional theory from this symmetry is obtained. The magnetic field interaction is seen to be equivalent to first order
to the interaction of a complex charged field with an external electric potential, where this 
ficticious "electric potential" is proportional to the external magnetic field. This allows one to solve
for the scattering amplitudes using already known scalar QED results. From the scalar QED analog the axion and
the photon are symmetric and antisymmetric combinations of particle and antiparticle. If one considers therefore
scattering experiments in which the two spatial dimensions of the effective theory are involved non trivially, one 
observes that both particle and antiparticle components of photons and axions are preferentially scattered in different 
directions, thus producing the splitting or decomposition of the photon and axion into their particle and antiparticle components
in an inhomogeneous magnetic field. This observable in principle effect is of first order in the axion photon coupling, 
unlike the  "light shining through a wall phenomena ", which is second order.
\end{abstract}


\maketitle

\section{Introduction}

The possible existence of a light pseudo scalar particle is a very interesting possibility. For example
the axion \ct{Peccei}, \ct{weinberg}, \ct{Wilczek}  which was introduced in order to solve the strong CP problem has
since then also been postulated as a candidate for the dark matter.
A great number of ideas and experiments for the search this particle have been proposed
\ct{Goldman}, \ct{Review}.

Here we are going to focus on a particular feature of the axion field $\phi$, which is its coupling to the photon 
through an interaction term of the form 
$g \phi \epsilon^{\mu \nu \alpha \beta}F_{\mu \nu} F_{\alpha \beta}$.
In fact a coupling of this sort is natural for any pseudoscalar interacting with
electromagnetism , as is the case of the neutral pion coupling to photons (which
as a consequence of this interaction decays into two photons).

A way to explore for observable consequences of the coupling of a light scalar 
to the photon in this way is to subject a beam of photons to a very strong magnetic field.

This affects the optical properties of light which could lead to testable consequences\ct{PVLAS},
also a magnetic field in the early universe can lead to interesting photon-axion conversion effects \ct{Yanagida} 
and in the laboratory photon-axion conversion effects could be
responsible for the "light shining through a wall phenomena ", which are is obtained by first
producing axions out of photons in a strong magnetic field region, then subjecting the mixed beam of photons
and axions to an absorbing wall for photons, but almost totally transparent to axions due to their weak 
interacting properties which can then
go through behind this "wall", applying then another magnetic field one can recover once again some photons 
from the produced axions \ct{LSTW}, \ct{Rabadan}. Notice however that the "light shining through a wall phenomena "
involves two interactions, once to produce the axions and then to obtain photons once again from the produced axions.
Since the axion photon coupling is so small, the amplitude for such effect is highly suppressed.

In this paper we will show that photons and axions splitt in the presence of an external magnetic field, in a way that we will make more precise. By this we mean that  from beam of photons we will get two different kind of scattered components (plus the photons that do not suffer any interactions), each of the scattered beams has also an axion component, but each of the beams
is directly observable due to its photon component and an observable process is obtained to first order in the axion photon 
interaction, unlike the "light shining through a wall phenomena ". Although we cannot claim yet that this could provide a more favorable 
experimental set up, since such subject involves many practical questions in addition to the existence of the first order process. 
Beyond the question of observability, the existence of this kind of effects highlights many basic features of the axion photon system.

\section{Action and Equations of Motion}
The action principle describing the relevant light pseudoscalar coupling to the photon is
\be
S =  \int d^{4}x 
\Bigl\lb -\frac{1}{4}F^{\mu\nu}F_{\mu\nu} + \frac{1}{2}\pa_{\mu}\phi \pa^{\mu}\phi - 
\frac{1}{2}m^{2}\phi^{2} - 
\frac{g}{8} \phi \epsilon^{\mu \nu \alpha \beta}F_{\mu \nu} F_{\alpha \beta}\Bigr\rb
\lab{axion photon ac}
\ee

We now specialize to the case where we consider an electromagnetic field with propagation along the y and z directions
and where a strong magnetic field pointing in the x-direction is present. This field may have an arbitrary space dependence in y and z, but it is assumed to be time independent. In the case the magnetic field is constant, see for example \ct{Ansoldi} for general solutions.

For the small perturbations we consider only small quadratic terms in the action for the axion fields and the electromagnetic field, following the method of for example Ref. \ct{Ansoldi}, but now considering a static magnetic field pointing in the x direction  having
arbitrary y and z dependence and specializing to y and z dependent electromagnetic field perturbations and axion fields. This means that the interaction between the background field , the axion and photon fields
reduces to
 
\be
S_I =  - \int d^{4}x 
\Bigl\lb \beta \phi E_x\Bigr\rb
\lab{axion photon int}
\ee

where $\beta = gB(y,z) $. Choosing the temporal gauge for the photon excitations and considering only the x-polarization for the electromagnetic waves, since only this polarization couples to the axion, we get the following 2+1 effective dimensional action
(A being the x-polarization of the photon, so that $E_x = -\pa_{t}A$)

\be
S_2 =  \int dydzdt 
\Bigl\lb  \frac{1}{2}\pa_{\mu}A \pa^{\mu}A+ \frac{1}{2}\pa_{\mu}\phi \pa^{\mu}\phi - 
\frac{1}{2}m^{2}\phi^{2} + \beta \phi \pa_{t} A
\Bigr\rb
\lab{2 action}
\ee

Since we consider only $A=A(t,y, z)$, $\phi =\phi(t,y,z)$, we have avoided the integration over $x$, for
the same reason, in \rf{2 action} $\mu$ runs over $t$, $y$ and $z$  only . This leads to the equations

\be
\pa_{\mu}\pa^{\mu}\phi + m^{2}\phi =  \beta \pa_{t} A
\lab{eq. ax}
\ee

\be
\pa_{\mu} \pa^{\mu}A = - \beta \pa_{t}\phi 
\lab{eq. photon}
\ee

As is well known, in temporal gauge, the action principle cannot reproduce the Gauss 
constraint (here with a charge density obtained from the axion photon coupling) and has
to be imposed as a complementary condition. However this 
constraint is automatically satisfied here just because of the type of dynamical reduction
employed and does not need to be considered  anymore.

\section{The Continuous Axion Photon Duality Symmetry and the Scalar QED analogy}
Without assuming any particular y and z-dependence for $\beta$, but still insisting that 
it will be static, we see that in the case $m=0$, we discover a continuous axion 
photon duality symmetry (these results were discussed previously in the 1+1 dimensional case, when only z dependence was considered in
\cite{duality}), since,

1. The kinetic terms of the photon and
axion allow for a rotational $O(2)$ symmetry in the axion-photon field space.

2. The interaction term, after dropping  a total time derivative can also be expressed in 
an $O(2)$ symmetric way as follows

\be
S_I =  \frac{1}{2} \int dydzdt 
\beta \Bigl\lb \phi \pa_{t} A - A \pa_{t}\phi \Bigr\rb
\lab{axion photon int2}
\ee

The axion photon symmetry is in the infinitesimal limit
\be
\delta A = \epsilon \phi, \delta \phi = - \epsilon A
\lab{axion photon symm}
\ee

where $\epsilon$ is a small number. Using Noether`s theorem, this leads to the 
conserved current $j_{\mu}$, with components given by

\be
j_{0} = A \pa_{t}\phi - \phi \pa_{t} A + \frac{\beta}{2}(A^{2} + \phi^{2} )
\lab{axion photon density}
\ee
and 
\be
j_{i} = A \pa_{i}\phi - \phi \pa_{i} A 
\lab{axion photon current}
\ee

Here $i= y, z$ coordinates . We define now the complex field $\psi$ as
\be
\psi = \frac{1}{\sqrt{2}}(\phi + iA)
\lab{axion photon complex}
\ee
we see that 
in terms of this complex field, the axion photon density takes the form
\be
j_{0} = i( \psi^{*}\pa_{t}\psi - \psi \pa_{t} \psi^{*}) +  \beta \psi^{*}\psi 
\lab{axion photon density complex}
\ee

We observe that to first order in $\beta$, \rf{axion photon int2} represents the
interaction of the magnetic field with the "axion photon density" \rf{axion photon density}, \rf{axion photon density complex} 
and also this interaction has the same form as that of scalar QED with an external "electric " field to first order.
In fact the magnetic field or more precisely $\beta /2$ appears to play the role of external electric potential 
that couples to the axion photon density \rf{axion photon density},\rf{axion photon density complex} which appears then to play the role
of an electric charge density . From this analogy one can obtain without effort the scattering amplitudes,
just using the known results from the scattering of charged scalar particles under the influence of an external
static electric potential, see for example \ct{Bjorken-Drell}.

One should notice however that the natural initial states used in a real experiment , like an initial photon and no axion involved, is not going to have a well defined axion photon charge in the second
quantized theory (although its average value appears zero), so the 
S matrix has to be presented in a different basis than that of normal QED . This is similar to the difference between working with linear polarizations as opposed to circular polarizations in ordinary optics, except that here we talk about polarizations in the axion photon space.
In fact pure axion and pure photon initial states correspond to symmetric and antisymmetric linear combinations of particle and antiparticle 
in the analog QED language. The reason these linear combinations are not going to be mantained in the presence on $B$  in the analog 
QED language, is that the analog external electric potential breaks the symmetry between particle and antiparticle and therefore will not
mantain in time the symmetric or antisymmetric combinations.

From the point of view of the axion-photon conversion experiments,
the symmetry \rf{axion photon symm} and its finite form , which is just
a rotation in the axion-photon space, implies a corresponding symmetry
of the axion-photon conversion amplitudes, for the limit $\omega >>m$.

In terms of the complex field, the axion photon current takes the form
\be
j_{k} = i( \psi^{*}\pa_{k}\psi - \psi \pa_{k} \psi^{*}) 
\lab{axion photon current complex}
\ee

\section{The Particle Anti-Particle Representation of Axions and Photons and their Splitting in an External Magnetic Field}
Now let us introduce the charge conjugation \cite{part-antipart} , that is,

\be
\psi \rightarrow \psi^{*}  
\lab{charge conjugation}
\ee

We see then that the free part of the action  is indeed invariant under \rf{charge conjugation}.
The $A$ and $\phi$ fields when acting on the free vacuum give rise to a photon and an axion respectivelly,
but in terms of the particles and antiparticles defined in terms of  $\psi$, we see that a photon is an antisymmetric 
combination of particle and antiparticle and an axion a symmetric combination, since

\be
\phi =\frac{1}{\sqrt{2}}(\psi^{*} +\psi), A= \frac{1}{i\sqrt{2}}(\psi - \psi^{*})  
\lab{part, antipart}
\ee

So that the axion is even under charge conjugation, while the photon is odd.
These two eigenstates of charge conjugation will propagate without mixing as long as no external magnetic field is applied.
The interaction with the extenal magnetic field is not invariant under \rf{charge conjugation}, in fact
under \rf{charge conjugation} we can see that
\be
S_I \rightarrow - S_I
\lab{non invariance}
\ee

Therefore  these symmetric and antisymmetric combinations, corresponding to axion and photon are not going to be mantained in the presence on $B$  in the analog QED language, since the "analog external electric potential" breaks the symmetry between particle and antiparticle and therefore will not mantain in time the symmetric or antisymmetric combinations.  In fact if the analog external electric potential is taken to be a
repulsive potential for particles, it will be an attractive potential for antiparticles, so the symmetry breaking is maximal.

Even at the classical level these two components suffer opposite forces, so both a photon or an axion under the influence of an inhomogeneous
magnetic field will be decomposed through scattering into their particle and antiparticle components, each of which is scattered in a different direction, since the analog electric force is related to the gradient of the effective electric potential, i.e., the gradient of the magnetic field, times the $U(1)$ charge which is opposite for particles and antiparticles.

For this effect to have meaning, we have to work at least in a 2+1 formalism, the 1+1 reduction \ct{duality}, \ct{part-antipart} which allows motion only in a single spacial direction is unable to produce such separation, since in order to separate particle and antiparticle components we need at least two dimensions
to obtain a final state with particles and antiparticles going in slightly different directions.

This is in a way similar to the Stern Gerlach experiment in atomic physics
\ct{Stern Gerlach}, where different spin orientations suffer a different force proportional to the gradient of the magnetic field in the direction of the spin. Here instead of spin we have that the photon is a combination of two states with different $U(1)$ charge and each of these components will suffer opposite force under the influence of the external inhomogeneous magnetic field. Notice also that since particle and antiparticles are distinguishable, there are no interference effect between the two processes.

Therefore an original beam of photons will be decomposed through scattering into two different elementary particle and antiparticle components
plus the photons that have not undergone scattering. These two beams are observable, since they have both photon components, so the observable
consequence of the axion photon coupling will be the splitting by a magnetic field of a photon beam. This effect being however an effect of first order in the axion photon coupling, unlike the  "light shining through a wall phenomena ".

\section{Conclusions}
The limit of zero axion mass when considering the scattering of axions and photons with the geometry relevant to the
axion-photon mixing experiments reveals a continuous axion photon duality symmetry.
This symmetry leads to a conserved current and then one observes that the interaction of the exteral magnetic field 
with the axion and photon is, to first order in the magnetic field, of the form of the first order in coupling constant 
interaction of charged scalars with an external electric scalar potential. Here the role of this ficticious external 
electric scalar potential is played (up to a constant) by the external magnetic field. 

Pure axion and pure photon initial states correspond to symmetric and antisymmetric linear combinations of particle and antiparticle 
in the analog QED language. Notice in this respect that charge conjugation of \rf{axion photon complex} corresponds to sign reversal of the photon field. The reason these linear combinations are not going to be mantained in the presence on a nontrivial $B$ in the analog 
QED language, is that the analog external electric potential breaks the symmetry between particle and antiparticle and therefore will not
mantain in time the symmetric or antisymmetric combinations.  

In this paper we present the 2+1 dimensional generalization of our previous work that allowed only 1+1 reductions \ct{duality}
\ct{part-antipart} . One possible application of this that has not been discussed here could be the generalization of the soliton solutions found in \ct{soliton}.

We have focused now on the implications of representing a photon (and also the axion) as a linear combination of particle and antparticles.
Even at the classical level these two components suffer opposite forces, so both a photon or an axion under the influence of an inhomogeneous
magnetic field (since the analog electric force is related to the gradient of the effective electric potential, i.e., the gradient of the magnetic field) will be decomposed through scattering into its particle and antiparticle components, each of which is scattered in a different direction. For this effect to have meaning, we have to work at least in a 2+1 formalism, since the 1+1 reduction \ct{duality}, \ct{part-antipart}, which allows motion only in a single spatial direction, is unable to describe such separation, since in order to separate particle and antiparticle components we need at least two dimensions (in order to obtain a final state with particles and antiparticles in slightly different directions). Notice also that since particle and antiparticles are distinguishable, there are no interference effect between the two processes.

Therefore an original beam of photons will be decomposed through scattering into two different elementary particle and antiparticle components
plus the photons that have not undergone any scattering. These two beams are observable, since they have both photon components, so the observable
consequence of the axion photon coupling will be the splitting by a magnetic field of a photon beam. This effect being however an effect of first order in the axion photon coupling, unlike the  "light shining through a wall phenomena ". In this process one should account for "decoherence"
of the particle and antiparticle once they are well separated. As we see, beyond the question of observability, the existence of this kind of effects highlights many basic features of the axion photon system.

This resembles the Stern Gerlach experiment in atomic physics
\ct{Stern Gerlach},  where different spin orientations suffer a different force proportional to the gradient of the magnetic field in the direction of the spin. Here instead of spin we have that the photon is a combination of two states with different $U(1)$ charge and each of these components will suffer opposite force under the influence of the external inhomogeneous magnetic field.

Our analysis will apply also to neutral pions and photons, for example a very energetic gamma ray scattering from an inhomogeneous magnetic field
could give rise to two scattered beams (each of them containing both pions and photons) if the scattering takes place in the plane orthogonal to the magnetic field. Possible observable effects of photon splitting out of cosmic magnetic fields and not just laboratory ones could also be considered as a new source of multiple images for example.

\section*{Acknowledgments}

I would like to thank the Physics Department of the City University of Osaka and in particular to Hideki Ishihara and Ken-Ichi Nakao for hospitality.

\end{document}